\documentclass[twocolumn, prl, showpacs, superscriptaddress]
{revtex4}%
\usepackage{amsmath}
\usepackage{color}
\usepackage{graphicx}
\usepackage{bm}

\begin{document}

\title[Coherence expansion in a microcavity]{Coherence expansion and polariton condensate formation in a semiconductor microcavity}

\author{V.V. Belykh}
\email{belykh@lebedev.ru} \affiliation{P.N. Lebedev Physical
Institute, Russian Academy of Sciences, Moscow, 119991 Russia}

\author{N.N. Sibeldin}
\affiliation{P.N. Lebedev Physical Institute, Russian Academy of
Sciences, Moscow, 119991 Russia}

\author{V.D. Kulakovskii}
\affiliation{Institute of Solid State Physics, Russian Academy of
Sciences, Chernogolovka, 142432 Russia}

\author{M.M. Glazov}
\affiliation{Ioffe Physical-Technical Institute of the Russian
Academy of Sciences, St. Petersburg, 194021 Russia}

\author{M.A. Semina}
\affiliation{Ioffe Physical-Technical Institute of the Russian
Academy of Sciences, St. Petersburg, 194021 Russia}

\author{C. Schneider}
\affiliation{Technische Physik, Physikalisches Institut and Wilhelm
Conrad R\"{o}ntgen Research Center for Complex Material Systems,
Universit\"{a}t W\"{u}rzburg, D-97074 W\"{u}rzburg, Germany}

\author{S. H\"{o}fling}
\affiliation{Technische Physik, Physikalisches Institut and Wilhelm
Conrad R\"{o}ntgen Research Center for Complex Material Systems,
Universit\"{a}t W\"{u}rzburg, D-97074 W\"{u}rzburg, Germany}

\author{M. Kamp}
\affiliation{Technische Physik, Physikalisches Institut and Wilhelm
Conrad R\"{o}ntgen Research Center for Complex Material Systems,
Universit\"{a}t W\"{u}rzburg, D-97074 W\"{u}rzburg, Germany}

\author{A. Forchel}
\affiliation{Technische Physik, Physikalisches Institut and Wilhelm
Conrad R\"{o}ntgen Research Center for Complex Material Systems,
Universit\"{a}t W\"{u}rzburg, D-97074 W\"{u}rzburg, Germany}

\begin{abstract}
The dynamics of the expansion of the first order spatial coherence
$g^{(1)}$ for a polariton system in a high-Q GaAs microcavity was
investigated on the basis of Young's double slit experiment under 3
ps pulse excitation at the conditions of polariton Bose-Einstein
condensation. It was found that in the process of condensate
formation the coherence expands with a constant velocity of about
$10^8$~cm/s. The measured coherence is smaller than that in
thermally equilibrium system during the growth of condensate density
and well exceeds it at the end of condensate decay. The onset of
spatial coherence is governed by polariton relaxation while
condensate amplitude and phase fluctuations are not suppressed.
\end{abstract}

\pacs{78.67.Pt, 71.36.+c, 78.47.jd, 03.75.Kk}

\maketitle

One of the most important characteristics of Bose-Einstein
condensate is the spatial coherence or the off-diagonal long range
order, i.e. the property of the system to share the same wave
function at different points separated by a distance larger than the
thermal de Broglie wavelength. To understand the processes governing
the Bose-Einstein condensation (BEC), it is important to know how
fast the coherence is established throughout the system during the
condensate formation. This question was addressed theoretically in
Refs. \cite{Kagan1, Kagan2}, where it was shown that in the process
of a condensation particles first relax to the low energy or
so-called coherent region where the kinetic energy of the particle
is of the order of its interaction energy with other particles.
Second, the fluctuations of density are smoothed out and a
``quasicondensate'' is formed. Further, the phase fluctuations
disappear resulting in the long range order formation within the
system signifying the onset of the ``true condensate''. The
timescales of these processes are quite different: the relaxation to
the low energy region is mainly determined by the stimulated
scattering processes, whereas the quasicondensate formation is
governed by the interparticle interactions. Experimentally, the
dynamics of spatial coherence formation was first studied for a gas
of ultracold atoms in Ref. \cite{Ritter1}, where it was found that
the coherence expands with a constant velocity of about 0.1 mm/s.

In this Letter we discuss the expansion of the spatial coherence in
a condensate of mixed exciton-photon states, polaritons, in a
semiconductor microcavity (MC) with embedded quantum wells. The
bosonic statistics of these particles and the extremely light
effective mass $m$ ($\sim 10^{-4}$ of the free electron mass $m_e$)
allow observing MC polariton BEC up to the room temperatures, which
inspired a considerable attention to this system in the last decade.
Up to now a number of bright phenomena in the MC polariton system
have been observed and discussed: polariton BEC \cite{Kasp1},
superfluidity \cite{Amo1}, quantized vortices \cite{Lag1},
spin-Meissner \cite{Lar1} and Josephson effect \cite{Lag2} (see
\cite{Sanvitto1} for a review). Compared with atomic BEC, the MC
polariton condensation is highly specific: the polariton system is
strongly nonequilibrium due to the short lifetime \cite{Tassone1}.

The first order spatial coherence function characterizes the ability
of the polariton system to interfere~\cite{Ons}, and is related to
off-diagonal elements of density matrix $\varrho(\bm \rho_1,\bm
\rho_2)$ in the coordinate space:
\begin{equation}
\label{def} g^{(1)}(\bm \rho_1 - \bm \rho_2) = \frac{\varrho(\bm
\rho_1,\bm \rho_2)}{\sqrt{\varrho(\bm \rho_1,\bm \rho_1)\varrho(\bm
\rho_2,\bm \rho_2)}}.
\end{equation}
The $g^{(1)}$ can be probed in the MC polariton system by measuring
the interference of the light emitted from different points on the
sample, since the amplitude and phase of the electric field of the
cavity emission is directly proportional to the amplitude and phase
of the wavefunction of the polariton condensate \cite{Kasp1, Deng1,
Szy}. Dynamics of $g^{(1)}$ in a process of MC polariton
condensation was studied experimentally for the first time in a
CdTe-based MC for the fixed separation between condensate regions
\cite{Nar1} and very recently in a GaAs-based MC \cite{Ohadi1}. In
the present work the spatial coherence dynamics of the polariton
system in a high-Q GaAs-based MC is studied in detail for different
separations $\Delta x$ between condensate regions. As a result it
was found for the first time that the coherence expands with almost
constant velocity and its value was measured. By tracing the
dependencies $g^{(1)}(\Delta x)$ at different times we extract the
dynamics of the coherence length $r_c$. We have found that in the
time range of the condensate decay, the coherence is larger than
that in the equilibrium system, which is highly unexpected. Our
study indicates that in a polariton system under pulsed excitation
of the MC the formation of the spatial coherence is governed by the
processes of polariton relaxation towards the ground state, whereas
amplitude and phase fluctuations of the quasicondensate are not
suppressed.

The sample is a half wavelength MC with Bragg reflectors made of 32
(for the top mirror) and 36 (for the bottom mirror) AlAs and
Al$_{0.13}$Ga$_{0.87}$As pairs. It has a Q-factor of about 7000 and
the Rabi splitting of 5 meV. The experiments were performed at
$T=10$~K and photon-exciton detuning of $-9$ meV. The sample was
excited by the radiation of a mode-locked Ti-sapphire laser
generating a periodic ($f=76$ MHz) train of 2.5-ps-long pulses at
the reflection minimum of the mirror 11 meV above the bare exciton
energy. The beam was focused in a 20~$\mu$m spot on the sample
surface. The spot was imaged with magnification of $\Gamma=6$ on the
light-absorbing plate with two transparent parallel slits
\cite{Deng1}. The interference pattern of the emission coming from
the regions of the sample selected by the two slits was formed on
the slit of the Hamamatsu streak camera, operating with time
resolution of 3~ps. Spatial coherence $g^{(1)}$ was extracted as the
visibility of the interference pattern
$g^{(1)}=(I_{max}-I_{min})/(I_{max}+I_{min})$, where $I_{min}$ and
$I_{max}$ are minimal and maximal intensities within one period of
interference pattern, averaged over all the observed periods. The
time-resolved MC emission spectra were recorded by a spectrometer
coupled to the streak camera with spectral (temporal) resolution of
0.25 meV (20 ps).

To convert the intensity $I(t)$, measured by the streak-camera to
the number of polaritons $N$ at states with wavevectors
$|k|<3$~$\mu$m$^{-1}$ (the collection aperture is $40^0$), the
integrated intensity of the MC photoluminescence (PL) $I_{PL}$ was
measured by the sensitive power meter. The number of polaritons was
evaluated by the relation $N(t)= 2 I_{PL}\tau_{LP}I(t)/(f
\hbar\omega \int I(t)dt)$, where factor 2 takes into account two
directions of photon emission, $\hbar\omega$ is the energy of
emitted photons, $\tau_{LP} \approx 3$~ps is the polariton lifetime
at the bottom of the lower polariton (LP) branch.
\begin{figure}
\begin{center}
\includegraphics[width=0.8\columnwidth]{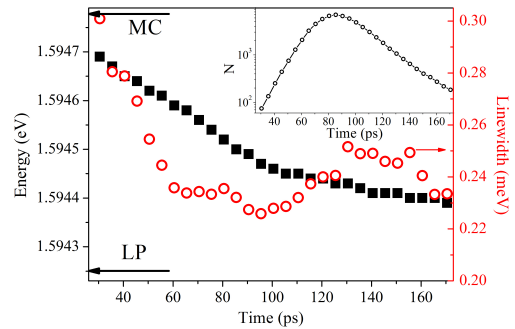}
\caption{Dynamics of the energy (full squares, left axis) and FWHM
(empty circles, right axis) of the LP emission line. $E_{MC}$ and
$E_{LP}$ are marked by black arrows. Inset shows the number of LPs
near $k=0$. Excitation power is $1.8 P_{thr}$.} \label{Fig-IEW}
\end{center}
\end{figure}

At low excitation density $P$ the PL dynamics of the LP branch is
relatively slow, and the angular distribution of intensity indicates
a bottleneck effect. As $P$ is increased above the BEC threshold
$P_{thr}=0.7$~kW/cm$^2$ (this value corresponds to the time-averaged
power of pulsed excitation), a fast and intense component in the PL
dynamics corresponding to $k \approx 0$ appears. Onset of the fast
component is accompanied by the blueshift of the spectral line and
decrease of its width (Fig.~\ref{Fig-IEW}). The energy position of
the spectral line is close the bare MC mode $E_{MC}$ just after the
excitation pulse and relaxes to the energy of LP mode $E_{LP}$ with
time. The maximum population (inset in Fig.~\ref{Fig-IEW}) is
reached when the spectral line energy is between the MC and LP modes
indicating that BEC is observed in the strong coupling regime
despite the high particle density \cite{Keel1, Kamide1, Byrnes1}.
\begin{figure*}
\begin{center}
\includegraphics[width=1.75 \columnwidth]{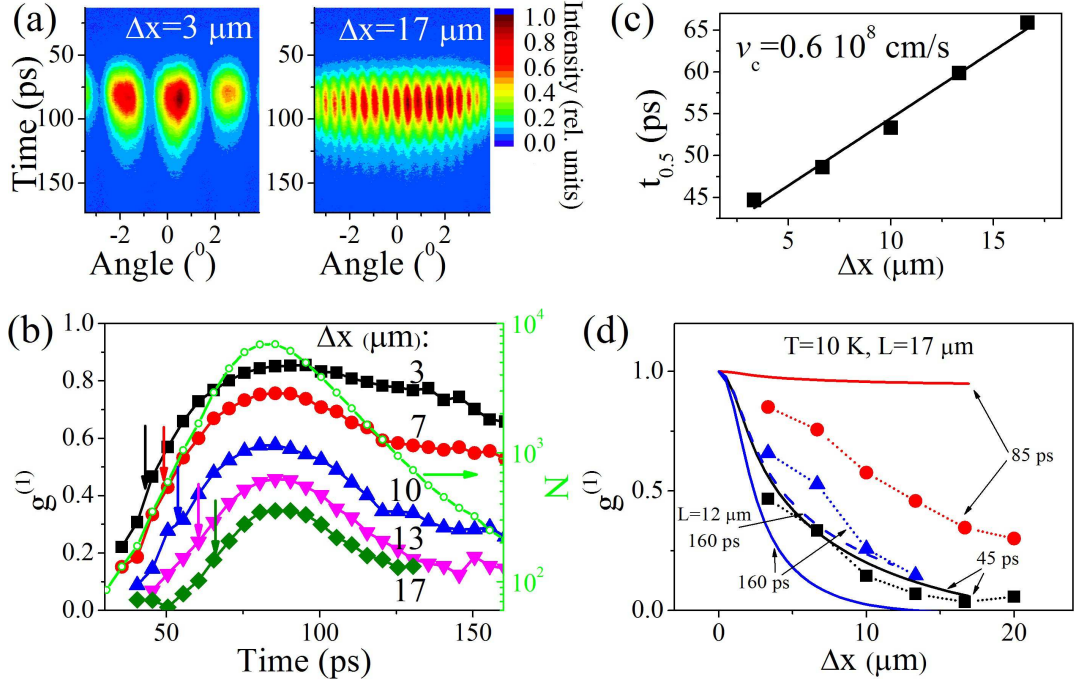}
\caption{(a) Streak camera images of the interference patterns of
the emission coming from regions of the sample separated by $\Delta
x = 3$ and 17~$\mu$m.  Horizontal axis corresponds to the angle of
the emission passed through the slits with respect to the sample
normal. (b) Dynamics of the spatial coherence for different $\Delta
x$ (left axis) and number of particles near the LP branch bottom
(right axis). Arrows mark the times $t_{0.5}$, when $g^{(1)}$
reaches half of its maximal value. (c) Dependence of the coherence
buildup time $t_{0.5}$ on $\Delta x$. Solid curve shows linear fit.
(d) Dependences $g^{(1)}(\Delta x)$ at different times $t$.
Experimental results are shown by symbols, solid lines show
dependences calculated at the same times as experimental ones for
the condensate size $L=17$~$\mu$m, and blue dashed line is
dependence calculated at $t=160$~ps and $L=12$~$\mu$m. Excitation
power for (a)-(d) is $1.8 P_{thr}$.} \label{Fig-g1}
\end{center}
\end{figure*}

Below $P_{thr}$ interference fringes in the double slit experiment
are observed only for the smallest studied slits separation, and
$g^{(1)}$($\Delta x =3$~$\mu$m)$<0.3$ at $0.9P_{thr}$. Above
$P_{thr}$ the interference fringes are well resolved up to $\Delta x
= 20$~$\mu$m (Fig.~\ref{Fig-g1}(a)).
The dynamics of $g^{(1)}$ for different $\Delta x$ at $P=1.8P_{thr}$
is presented in Fig.~\ref{Fig-g1}(b) together with the dynamics of
the polariton number $N$ at the bottom of the LP branch.
Figure~\ref{Fig-g1}(b) shows that the maximum value of $g^{(1)}$
decreases with increased $\Delta x$ and the decay of $g^{(1)}$ in
the whole range of $\Delta x$ occurs much slower than that of the
condensate density in agreement with \cite{Nar1}.

Interestingly, Fig.~\ref{Fig-g1}(b) shows that the coherence buildup
time increases with an increase of $\Delta x$, indicating a finite
velocity of the coherence expansion. We define the coherence buildup
time $t_{0.5}(\Delta x)$ as the time when $g^{(1)}$ reaches half of
its maximum value for a given $\Delta x$. These times are marked by
arrows in Fig.~\ref{Fig-g1}(b). Figure~\ref{Fig-g1}(c) shows that
the dependence of $t_{0.5}$ on $\Delta x$ is close to linear
indicating that the coherence expands with almost constant velocity
$v_c=0.6\cdot10^8$~cm/s.

\begin{figure}
\includegraphics[width=\columnwidth]{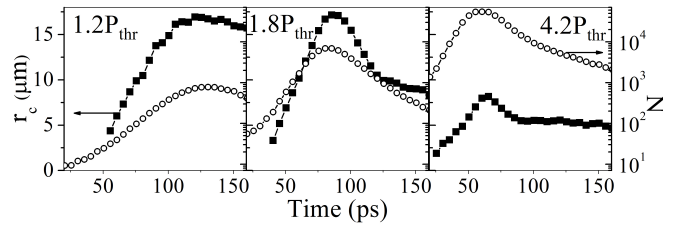}
\caption{Dynamics of the coherence length (full squares, left axes,
linear scales) and number of particles near the bottom of the LP
branch (empty circles, right axes, logarithmic scales) at different
excitation powers.} \label{Fig-rc}
\end{figure}

Figure~\ref{Fig-rc} shows the dynamics of the coherence length
$r_c$, which is defined from $g^{(1)}(r_c)=1/e$, and number of
polaritons at the bottom of the LP branch $N$ for $P>P_{thr}$. At
the beginning of the BEC, $r_c$ grows almost linearly with time,
reaches its maximum and decays afterwards. It follows from
Fig.~\ref{Fig-rc} that $r_c$ and $N$ reach their maximal values
almost at the same time, but the decay of $N$ at high $P$ occurs
much faster than that of $r_c$: $N$ decays with the lifetime of
20-30~ps whereas $r_c$ decreases at most by $50 \%$ during the first
30-40~ps and then remains nearly constant during the next several
tens of picoseconds where $N$ decreases by more than an order of
magnitude. Furthermore, Fig.~\ref{Fig-rc} shows that the buildup of
the coherence at $P=1.2P_{thr}$ begins when the particle number $N$
is more than one order of magnitude smaller than that at $P=4.2
P_{thr}$ and that the maximal value of $r_c$ decreases with $P$ at
$P>2 P_{thr}$ in spite of a strong (about an two orders of
magnitude) increase of the condensate density. These facts show that
the length of coherence, as well as the rate of its increase, is not
solely defined by the condensate occupation because of the
nonequilibrium nature of the polariton BEC.

In a 2D system of a finite size $L$ under the conditions of thermal
equilibrium the number of particles in the ground state can be
estimated as
\begin{equation}
N_0=N-N'=N-\int_{1/L}^{\infty} \frac{k dk L^2}{\pi}(e^\frac{\hbar^2
k^2}{2m k_B T}-1)^{-1},  \label{N0}
\end{equation}
where $N$ is the total number of LPs, $N'$ is the number of LPs in
all states but the ground, $k_B$ is the Boltzmann constant. Here the
chemical potential $\mu=0$.

It follows from Figs.~\ref{Fig-g1}(b) and \ref{Fig-rc} that the
onset of spatial coherence at $P < 2 P_{thr}$ starts at $N \sim
10^2$. This value is less than the estimated from Eq.~(\ref{N0}) $N'
\approx 400$ at $T=10$~K in the investigated LP system with the
lateral size $L\approx 17$~$\mu$m and $m=5 \cdot 10^{-5} m_e$. Thus,
the formation of spatial coherence starts at negative chemical
potential $\mu$ with respect to the bottom of the LP branch (e.g.
$\mu \sim -0.05$~meV for $P=1.8P_{thr}$, $t=40$~ps) and hence
governed by the relaxation process of polaritons to the low energy
region, while interaction-induced suppression of amplitude and phase
fluctuations plays no role \cite{Kagan1, Kagan2}. We note also, that
at $P<2P_{thr}$ the estimated value of the interaction energy for
the condensed polaritons $\Delta E = \alpha N_0/L^2<2$~$\mu$eV $\ll
k_B T \approx 1$~meV (where $\alpha=10^{-12}$~meVcm$^2$ is the
polariton-polariton interaction constant). Hence, $g^{(1)}$ should
be close to that for the classical noninteracting gas with the
particle distribution function $N_k$ (see Eq.~\eqref{def} and
\cite{Deng1}):
\begin{equation}
g^{(1)}(\Delta x) =\frac{\sum_{\bm k} N_k \mathrm e^{\mathrm i k
\Delta x}}{\sum_{\bm k} N_k}=
\frac{\frac{L^2}{\pi}\int^{\infty}_{1/L}J_0(k \Delta x)N_k k
dk+N_0}{\frac{L^2}{\pi}\int^{\infty}_{1/L}N_k k dk+N_0},
\label{g1:class}
\end{equation}
where $J_0(x)$ is the zero order Bessel function.

The detailed modeling of the polariton distribution similar to that
in Refs. \cite{Kin} is beyond the scope of the present paper. Here
we discuss the reason of the difference between the experimental
dependences $g^{(1)}(\Delta x)$ and calculated ones for the thermal
Bose distribution of LPs with the use of $T=10$~K and $\mu$
determined from the measured polariton number $N$ in the low energy
region. The difference between the calculated and measured curves
indicates how far the system is from the thermal equilibrium.

The results of the calculation for three consecutive stages:
condensation onset, maximum condensate density and its decay, are
shown in Fig.~\ref{Fig-g1}(d) for $P=1.8 P_{thr}$ and the condensate
size $L=17$~$\mu$m determined from the experiment. It is seen that
the experimental dependence $g^{(1)}(\Delta x)$ is slightly below
the calculated thermally equilibrium one in the first stage
($t=45$~ps), the difference strongly increases in the second stage
($t=85$~ps) whereas in the third stage ($t=160$~ps) experimental
values $g^{(1)}(\Delta x)$ turn out well above the calculated ones.
This result is especially surprising as it indicates that the
occupation of  $k \approx 0$ states with respect to that of
higher-energy states exceeds the thermally equilibrium one, i.e. the
effective polariton temperature is lower than the bath temperature
in the time range of condensate decay.
Measurements of polariton distribution along the LP branch (not
shown) indicate that the distribution function approaches the
thermal one at the onset of condensation, but the occupation numbers
of low $k \neq 0$ states are slightly increased compared to the
thermal values, which explains a small discrepancy between the
experimental and calculated $g^{(1)}(\Delta x)$ at $t=45$~ps, when
$N \approx 350 < N'$.

An increase in the discrepancy between the experimental and
calculated $g^{(1)}(\Delta x)$ in the range of maximal $N$ at $t=85$
ps when $N \approx 7000 \gg N'$ indicates that the LP system becomes
more nonequilibrium at high condensate density. The most probable
reason for that is the runaway of condensed polaritons from the
small bounded ($\sim 20$~$\mu$m) photoexcited region due to their
repulsive interaction with a dense exciton reservoir
\cite{Brichkin}. In addition, for the equilibrium system with $N_0
\gg N'$, the coherence is defined by the amplitude and phase
fluctuations of the ground state wavefunction not taken into account
in Eq.~(\ref{g1:class}).
Indeed, the quasicondensate amplitude fluctuations are suppressed
during the time $\tau_A \approx \hbar L^2 / (\alpha
N)$~\cite{Kagan1, Kagan2}, determined by the interparticle
interaction. For $N=7000$, $L=17$~$\mu$m we obtain value $\tau_A
\sim 0.3$~ns that exceeds the lifetime of LP condensate at $P=1.8
P_{thr}$ (Fig.~\ref{Fig-rc}). It follows then, that the amplitude
and, especially, phase fluctuations are not suppressed.

Finally, let us discuss the reason for the unexpectedly high
coherence in the decaying polariton system at $t \sim 160$~ps with
$N \approx 200 <N'$ which exceeds markedly the coherence in the
thermally equilibrium system. Note that the calculations
underestimate experimental values of $g^{(1)}$ even for the
unrealistically small condensate size $L=12$~$\mu$m (dashed line in
Fig.~\ref{Fig-g1}d). At large $t > 100$~ps the reservoir becomes
highly depleted, and the equilibrium in the system is established
mainly via ineffective excitation of condensed polaritons by
acoustic phonons with characteristic time exceeding the condensate
decay time $\tau$. In this case the wave function at a given
distance $\bm\rho$ decays as $\psi(\bm \rho)\propto
\exp{(-\frac{t}{2\tau})}$. Thus, both diagonal and off-diagonal
elements of the polariton density matrix decay approximately with
the same rate $1/\tau$ resulting in a nearly constant ratio of the
LP numbers in the ground state to that in the excited states well
exceeding the equilibrium ratio at long times. As a result,
$g^{(1)}$ weakly decays with time in agreement with the experiment.

It is worth to compare the measured coherence expansion velocity
$v_c$ for the polariton condensate (about $10^8$~cm/s) and that for
the atomic condensate (about $10^{-2}$~cm/s \cite{Ritter1}). The
buildup of the spatial coherence is determined by the relaxation of
the particles to the ground state and manifests itself by the
condition $r_c > \lambda_{dB}$, where $\lambda_{dB}$ is thermal de
Broglie wavelength. So one can estimate the coherence expansion
velocity as $v_c \sim \lambda_{dB} / \tau_{rel}$. For polaritons
$\lambda_{dB} \sim 2$~$\mu$m at $T=10$~K and for atoms $\sim
0.4$~$\mu$m at $T=0.2$~$\mu$K, so the large difference in $v_c$ is
related to the difference in the relaxation times $\tau_{rel}$,
which is $\sim 10$~ps for polaritons and $\sim 100$~ms for atoms
\cite{Ritter1}. The relaxation for both polaritons and atoms is
accomplished via interparticle scattering, while for polaritons
scattering with phonons also plays role. The rate of the
interparticle collisions depends on the scattering cross sections
and on the average velocity of the particles. The latter is
determined by the particle mass and temperature which differ in many
orders of magnitude ensuring the necessary ratio of the relaxation
times for polaritons and atoms.

To conclude, we have studied the dynamics of the spatial coherence
for a LP condensate under pulsed ps-long excitation for different
excitation powers.  It has been found that in the process of
condensate formation, first order coherence expands with almost
constant velocity of about $10^8$~cm/s. We have shown that the
coherence is influenced by polariton relaxation from the reservoir.
The onset of spatial coherence is determined by the narrowing of
polariton distribution in $k$-space rather than formation of the
condensate phase, and at high excitation density coherence is
limited by condensate amplitude fluctuations. The true condensate
i.e. macroscopic occupation of the ground state with suppressed
phase fluctuation is not achieved under ps-long pulsed pumping.

\begin{acknowledgements}
We are grateful to D.A. Mylnikov for help in the experiment and S.S.
Gavrilov, N.N. Gippius, L.V. Keldysh, A.V. Sekretenko, S.G.
Tikhodeev and V.B. Timofeev for valuable advice and useful
discussions. This study was supported by the RFBR (projects no.
11-02-01310, 11-02-12261, 11-02-00573), RAS, Ministry of Education
and Science of the Russian Federation (contract no. 8680), the State
of Bavaria, and EU projects POLAPHEN and SPANGL4Q. MAS was partially
supported by the RF President Grant NSh-2901.2012.2.
\end{acknowledgements}

\end{document}